\documentclass{neutralclass}

\usepackage{amsmath,amssymb}
\usepackage{dsfont}
\usepackage{subcaption}
\usepackage{enumitem}

\newcommand{\R}{\mathbb{R}}

\newcommand{\1}{\mathds{1}}

\newcommand{\X}{\mathcal{X}}
\newcommand{\x}{\mathbf{x}}

\renewcommand{\hat}{\widehat}
\renewcommand{\tilde}{\widetilde}
\renewcommand{\d}{\ensuremath {\,\mathrm{d}}}

\begin{document}

\title{Adjusting for Social Groups in Spatial Point Process Models for Waiting Pedestrian Configurations}
\titlerunning{SPPs for Waiting Pedestrians with Groups}

\author{
  Lars Sickert Karam\authorlabel{1} \and 
  Rui M. Castro\authorlabel{1} \and
  Maarten Schoukens\authorlabel{2} \and 
  Alessandro Corbetta\authorlabel{3}
}
\authorrunning{L. Sickert Karam et al.}
\institute{
  \authorlabel{1} Department of Mathematics \& Computer Science and Eindhoven Artificial Intelligence Systems Institute (EAISI), Eindhoven University of Technology, Eindhoven, The Netherlands,
  \authoremail{1}{l.sickert.karam@tue.nl, r.m.pires.da.silva.castro@tue.nl}
  \and
  \authorlabel{2} Department of Electrical Engineering and EAISI, Eindhoven University of Technology, Eindhoven, The Netherlands,
  \authoremail{2}{m.schoukens@tue.nl}
  \and
  \authorlabel{3} Department of Applied Physics \& Science Education and EAISI, Eindhoven University of Technology, Eindhoven, The Netherlands,
  \authoremail{3}{a.corbetta@tue.nl}
}

\maketitle

\begin{abstract}

The spatial distribution of waiting pedestrians has two primary drivers: environmental preference and interaction with other pedestrians. Spatial point processes naturally capture both spatial heterogeneity and repulsive interaction. In previous work (Sickert Karam et al., arXiv:2606.14532, 2026), we proposed a Gibbs model with inhomogeneous intensity and a modified Diggle–Gates–Stibbard interaction function, which reproduces many phenomena in replicated patterns of pedestrians waiting at a train station. Its single interaction function acts as an effective interaction, averaging over behavioral regimes such as interactions within social groups and among strangers. In this article, we show how groups affect distance-based summary statistics and take first steps towards accounting for them. We detect groups from pairwise distances and contact durations, replace each group by a ``macro-pedestrian'' at its average position, and refit the model. This yields a larger interaction range and better agreement in nearest-neighbor statistics than the original model, although the two fits concern different datasets and centroid-based distances likely overstate repulsion. This first attempt thus resolves some discrepancies but falls short of a fully adequate solution, opening avenues for further research.
\end{abstract}


\begin{figure}[t]
  \centering
  \includegraphics[width=\textwidth]{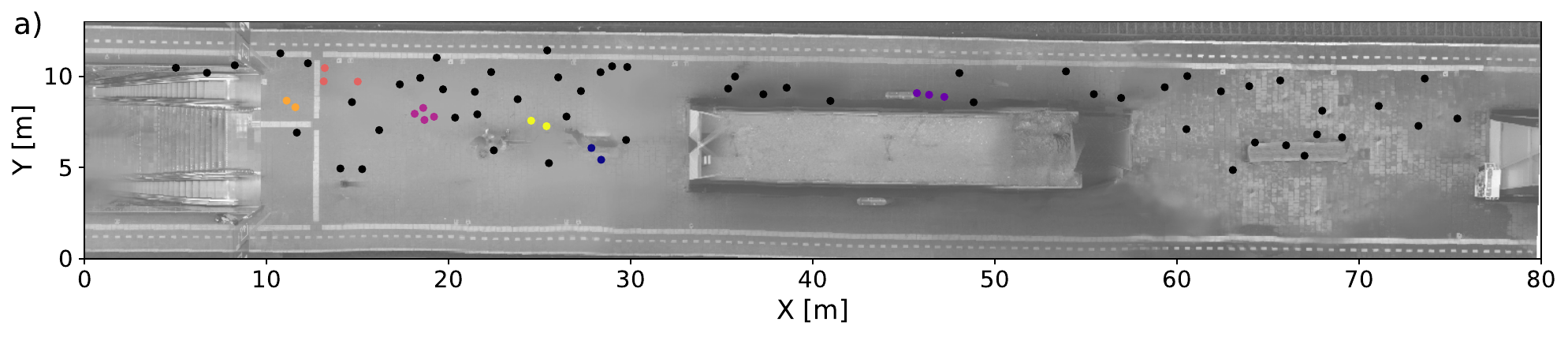}
  \includegraphics[width=\textwidth]{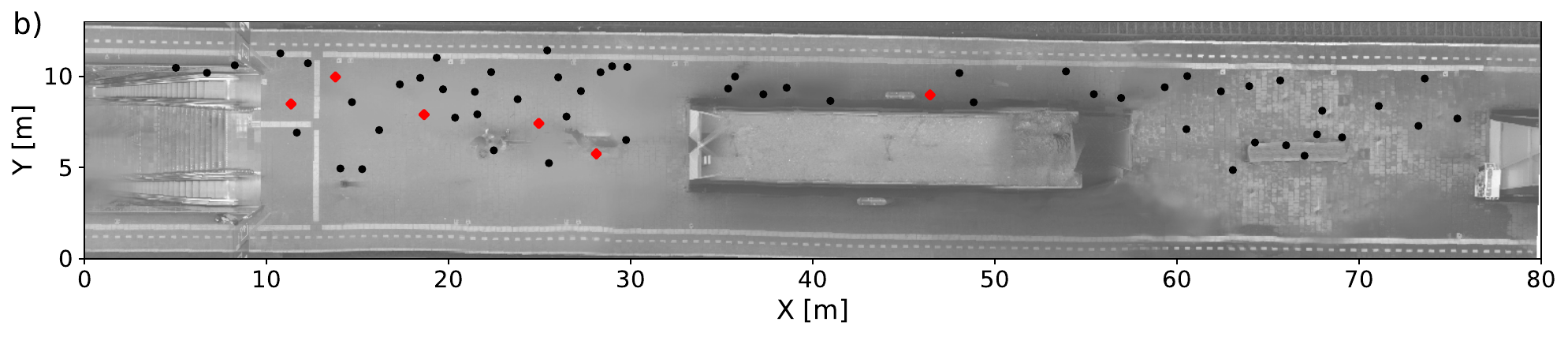}
  \caption{(a) Example snapshot with detected groups in color. Only waiting pedestrians are shown after filtering out passing pedestrians. (b) Same snapshot after aggregation, showing macro-pedestrians in red with new, averaged position.}
  \label{fig:snapshot}
\end{figure}

\section{Introduction}
\label{sec:intro}

Understanding how pedestrians wait on train station platforms, and modeling the spatial patterns formed by their positions, is highly relevant for the design and operation of transport infrastructure \cite{Kupper_2023a}. These patterns are driven primarily by two aspects: environmental preferences and interactions between pedestrians. Abstracting the pedestrians as points in the Euclidean plane leads to a natural modeling choice for these arrangements as a realization of a spatial point process \cite{Sickert_2026}, a stochastic model that can capture both background effects and interactions \cite{Baddeley_2015}. Waiting pedestrians exhibit repulsive interactions, as on the one hand, the distance between individuals is inherently limited by their physical size, and on the other hand, pedestrians may additionally prefer to keep a social distance from one another \cite{Helbing_1995}. In previous work \cite{Sickert_2026}, we proposed a Gibbs point process model \cite{vanLieshout_2000} with inhomogeneous intensity and a modified Diggle-Gates-Stibbard interaction function \cite{Sickert_2026,Diggle_1987}, which quantitatively reproduces many features of waiting crowds. In this model, all pairwise interactions are described by a single interaction function, which can be interpreted as an effective interaction averaging over different behavioral regimes present in a crowd. Such regimes include interactions between strangers and those between members of social groups, which commonly occur in pedestrian crowds \cite{Kupper_2023b}, so that short distances between individuals may reflect either weak global repulsion or membership of the same group. We build on this framework in the present work by explicitly accounting for the presence of groups and proposing first steps toward addressing their influence on the interaction structure of waiting pedestrians.

We begin by illustrating how groups affect distance-based summary statistics, and introduce a relatively simple correction based on the concept of ``macro-pedestrians'' -- replacing each group by a point at its average position (see \fref{fig:snapshot} for an example). Then, the Gibbs model is refit to the resulting patterns. This approach substantially increases the interaction range of the fitted model, while making the transition between interaction regimes (from full exclusion to independence) smoother. Moreover, simulations show that this new model fits the macro-pedestrian patterns better in terms of nearest-neighbor statistics than the original model \cite{Sickert_2026} fits the patterns of individual pedestrians. This comparison should be interpreted with care, however, as the two models are fitted to and evaluated on different datasets: after correcting for groups, the macro-pedestrian patterns contain fewer points and no longer include distances within groups. Furthermore, we argue that reducing groups to their centroid introduces new biases, as distances between singletons and groups are measured from group centers rather than their edges, favoring long ranges of interaction. This simple correction thus resolves some discrepancies but falls short of being a full-fledged solution, opening avenues for further research.

This paper is organized as follows. \sref{sec:data} describes the dataset, consisting of replicated spatial patterns (i.e. arrangements of points hypothesized to be generated by the same underlying process) of pedestrian positions collected at Eindhoven central station in the Netherlands. Additionally, the group detection methodology is discussed. In \sref{sec:model}, the Gibbs model is introduced, as well as the methodology used for estimation of the underlying inhomogeneous background intensity and the interaction parameters. In \sref{sec:results}, results of estimation under consideration of groups are presented and contrasted with results obtained from the model ignoring groups. A discussion in \sref{sec:discussion} closes the work.

\section{Waiting Pedestrian Data, Group Detection, and Aggregation}
\label{sec:data}

In this section, we present the dataset used for later analysis. It consists of configurations of waiting pedestrians at a train station, each a static snapshot in time. These snapshots can be viewed as replicated spatial patterns \cite{Baddeley_2015}, approximately independent realizations of a spatial point process. Crucially, this setting incorporates the combination of an inhomogeneous background intensity, i.e. attractiveness of different positions, and an interaction between points. This creates a unique statistical challenge, as it is inherently difficult to separate these two influences during estimation \cite{Sickert_2026,Baddeley_2015}.

\begin{figure}
  \centering
  \includegraphics[width=\textwidth]{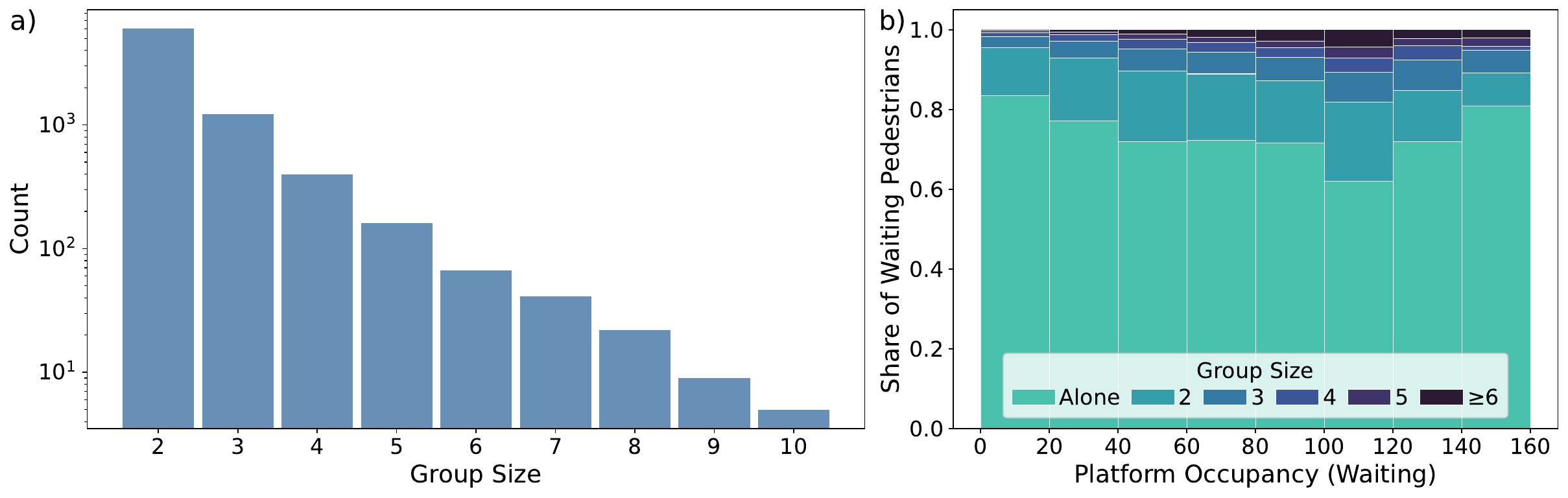}

  \caption{For both panels, data is restricted to snapshots with no groups larger than $10$ for visualization purposes. (a) Frequency of different group sizes (disregarding individuals); counts in log scale exhibit an exponential trend. (b) Share of waiting pedestrians by group size, per platform occupancy bin. Slight trend towards larger groups as occupancy increases.}
  \label{fig:histograms}
\end{figure}

As a baseline, we use the dataset of waiting pedestrian configurations introduced in \cite{Sickert_2026}, extracted from overhead sensor recordings on Platform $2$ of Eindhoven Central Station in the Netherlands. This data is publicly available (with only minor discrepancies, described in \cite{Sickert_2026}) at \cite{Sickert_2026b}. Pedestrian trajectories are recorded at $10\,$Hz and classified as \emph{waiting} or \emph{passing} using a velocity-based criterion. Configurations of waiting individuals are sampled one minute before peaks in platform occupancy, yielding replicated spatial patterns of high-occupancy scenarios while reducing the direct influence of train arrivals. The platform exhibits spatial heterogeneity due to entrances, obstacles, buildings, and seating areas. See \cite{Sickert_2026} for a more detailed description of the geometry and preprocessing.

At this point, we deviate from the dataset of \cite{Sickert_2026} (we will refer to this as the \emph{original dataset}) by considering social groups. We use the graph-based method proposed in \cite{Pouw_2020} and validated in \cite{Kupper_2023b} and \cite{vanderLaan_2026} to detect whether pedestrians belong to a social group. The group-detection algorithm identifies pairs of pedestrians that remain close to each other for a large fraction of the time during which both are observed, and then takes connected components of the resulting contact graph as social groups. For each pair of pedestrians, the time $t_{\text{contact}}$ for which the pedestrians are ``in contact,'' i.e. the number of frames where the distance is below a certain threshold (we choose $1.5\,$m, aligning with \cite{Pouw_2020,Kupper_2023b}), and the time $t_{\text{personal}}$ for which pedestrians are within ``personal distance,'' set at $1\,$m, are recorded. Pairs are then checked for two relations within a timeframe of $30\,$s\footnote{
  While longer timescales of $60\,$s and $300\,$s were investigated on subsets of the data, the groups they detected matched those detected at $30\,$s, while being computationally much more expensive.
} leading up to time of the snapshot being captured:

\begin{equation*}
  t_{\text{contact}} \geq a \cdot t_{\text{simultaneous}}; \quad t_{\text{personal}} \geq b \cdot t_{\text{simultaneous}},
\end{equation*}

where $t_{\text{simultaneous}}$ is the time in which both pedestrians are simultaneously in frame. If both conditions hold for predetermined values of $a, b$, the pair is determined to belong to the same social group. We choose $a=0.85$ and $b=0.4$ as proposed and validated in \cite{Kupper_2023b}, as the scenario tested there closely resembles our own. See \fref{fig:histograms} for some summary statistics of groups in the dataset, showing small group sizes are substantially more common than large ones, and how group size is related to platform occupancy. Snapshots containing groups larger than $5$ members are then excluded for further analysis, as their centroid may lie far away from the positions of the pedestrians at the edge of the group. We introduce the notation $N_{\text{groupmax}}$ to denote the maximum size of a group in the dataset, i.e. here $N_{\text{groupmax}} = 5$. This can artificially inflate nearest-neighbor distances in the full pattern, disproportionately affecting distance-based diagnostics later on. This filtering leaves us with $2457$ snapshots, compared to $2629$ snapshots in the original dataset of \cite{Sickert_2026}, see \fref{fig:snapshot}(a) for an example before aggregation.  Each detected group present in a snapshot is then represented by its centroid, while singleton pedestrians are left unchanged, leaving us with a dataset consisting of individuals and macro-pedestrians representing groups. See \fref{fig:snapshot}(b) for the same snapshot as panel (a) after aggregation.

\begin{figure}[t]
  \centering
  \includegraphics[width=\textwidth]{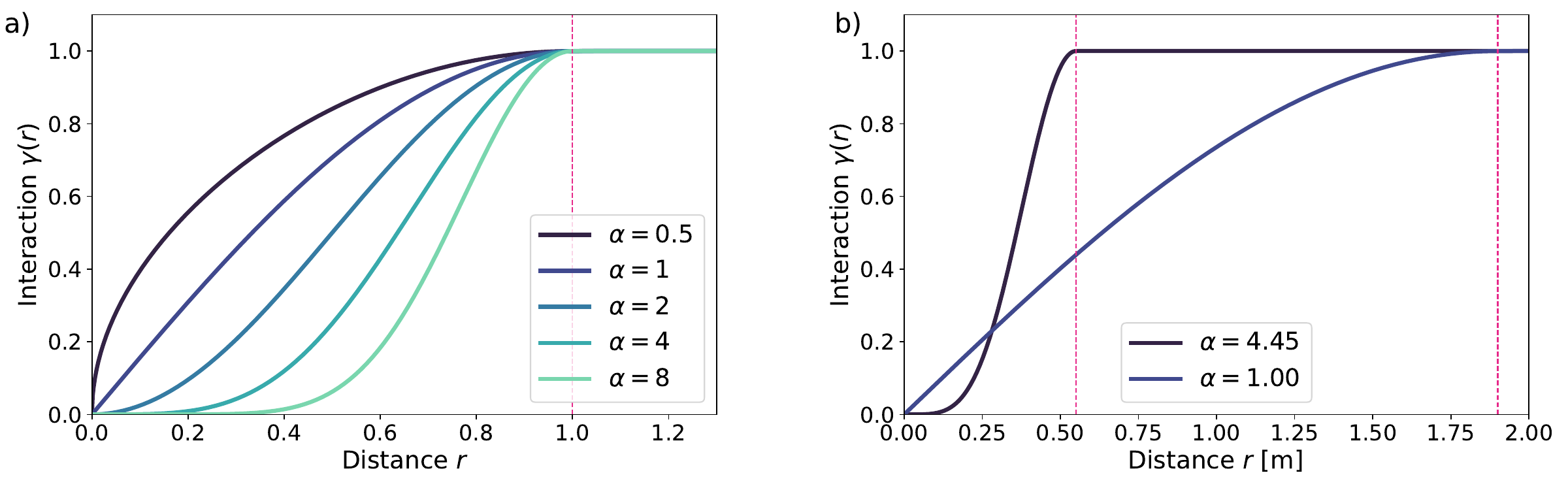}
  \caption{(a) DGS interaction function for multiple values of $\alpha$, illustrating the effect of the shape parameter. (b) DGS interaction functions with estimated parameters without ($R=550, \alpha=4.45$) and with ($R=1900, \alpha=1$) considering groups.}
  \label{fig:interactions}
\end{figure}

\section{Gibbs Point Processes for Waiting Pedestrians: Formal Details and Statistical Inference}
\label{sec:model}

In this section, to ensure this article is self-contained, we introduce Gibbs point processes and define the modified Diggle-Gates-Stibbard (DGS) interaction function, following \cite{Sickert_2026}. We then briefly review maximum pseudolikelihood estimation, conditional on the number of points in the configurations. The DGS interaction function is chosen pragmatically based on its performance on the dataset in \cite{Sickert_2026}, where it was best among tested models. It represents a radial repulsion that smoothly transitions from full exclusion to independence in a finite radius (see \fref{fig:interactions}).

Denote an (unordered) configuration of $n$ points in a connected subset $\X \subset \R^2$ by
\begin{equation*}
  \x = \{x_1, \dots, x_n\}, \quad x_i \in \X \text{ for all } i \in \{1, \dots, n\}.
\end{equation*}
Defining as $\X_n^*$ the space of configurations $\x \subset \X$ of size $N(\x) = n$, the probability density function (pdf) for a Gibbs point process with pairwise interaction conditional on the configuration size is
\begin{equation}\label{eq:gibbs_density}
  f(\x | N(\x) = n) = Z^{-1}(n, \beta, \gamma) \prod_{i=1}^n \beta(x_i) \prod_{\substack{i, j \in \{1, \dots, n\} \\ i < j}} \gamma(x_i, x_j),
\end{equation}
where $Z$ is a normalizing constant given by 
\begin{equation*}
  Z(n, \beta, \gamma) = \int_{\X} \cdots \int_{\X} \prod_{i=1}^n \beta(u_i) \prod_{\substack{i, j \in \{1, \dots, n\} \\ i < j }} \gamma(u_i, u_j) \d u_1 \cdots \d u_n.
\end{equation*}
Here, $\beta \colon \X \to \R_{\geq 0}$ is the \emph{intensity function}, a first order trend or spatial attractiveness function, and $\gamma \colon \X \times \X \to \R_{\geq 0}$ is the (pairwise) \emph{interaction function}. A value of $\gamma \equiv 1$ corresponds to no interaction, while values less than or greater than one correspond to repulsive or attractive interactions, respectively.

In this article, we consider a modified version of the \emph{Diggle-Gates-Stibbard} interaction function \cite{Sickert_2026, Diggle_1987}, which is a smooth interpolation between $0$ and $1$ in a range given by an interaction radius $R>0$ and a shape parameter $\alpha \geq 0$. It is defined as 
\begin{equation*}
    \gamma_\theta(x, y) = 
    \begin{cases}
    \Biggl( \sin \Bigl( \dfrac{\pi \, \|x - y\|}{2 R} \Bigr) \Biggr)^{2\alpha}, & \text{if } \|x - y\| < R, \\[2mm]
    1, & \text{if } \|x - y\| \geq R.
    \end{cases}
\end{equation*}
Larger values of $\alpha$ impose stronger repulsion at short ranges and a steeper transition toward no interaction as distances approach $R$, see Figure~\ref{fig:interactions} for an illustration.

Since $Z$ in \eref{eq:gibbs_density} is generally intractable, it is typically approximated. While Markov Chain Monte Carlo (MCMC) methods are available, they are computationally very expensive \cite{Geyer_1994}. A substantially less expensive approach more commonly used in practice is based on Besag's pseudolikelihood \cite{Besag_1975}. Essentially, in the conditional setting, this method approximates the normalizing constant in the likelihood, which is an $n$-fold product integral, by a product of $n$ one-dimensional integrals \cite{Sickert_2026,Moller_2022}. This results in the following form of the conditional log-pseudolikelihood for an interaction parameter vector $\theta = (\alpha, R)$:
\begin{equation*}
  \text{CPL}_n(\theta|\x) = \sum_{i=1}^n \left( \log \beta(x_i) + \sum_{y \in \x \backslash \{x_i\}} \log \gamma_\theta(x_i, y) - \log \tilde{Z}(\x \backslash \{x_i\}; \theta) \right),
\end{equation*}
with
\begin{equation*}
  \tilde{Z}(\x \backslash \{x_i\}; \theta) = \int_{\X} \beta(u) \prod_{\substack{t \in \x \backslash \{x_i\}}} \gamma_\theta(u, t) \d u.
\end{equation*}

\noindent We use the semiparametric estimation algorithm of \cite{Sickert_2026}. In brief, an initial kernel estimate of the intensity function $\beta$ is obtained under a Poisson assumption (i.e. no interaction between points), on the basis of which interaction parameters $\theta$ are estimated by conditional maximum pseudolikelihood. Having obtained parameters, the bandwidth for the kernel estimate of $\beta$ is modified by likelihood cross-validation over a set of rescalings of the original bandwidth. Crucially, the likelihood used for the cross-validation is the Gibbs likelihood with the parameter estimates for the interaction function plugged in. The final model is obtained by re-estimating the intensity function and maximizing the conditional pseudolikelihood again using the new estimate. This procedure can be iterated.

\begin{figure}[t]
  \centering

  \includegraphics[width=\textwidth]{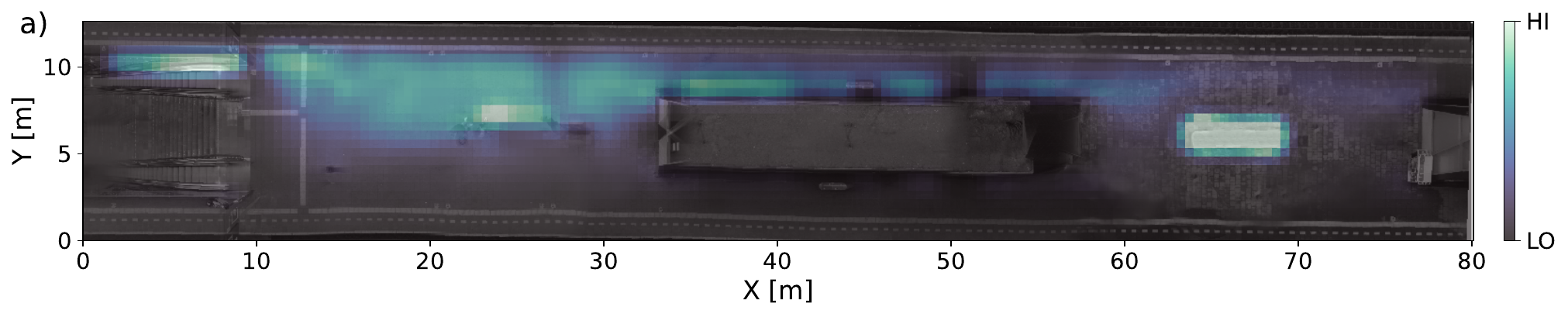}
  \includegraphics[width=\textwidth]{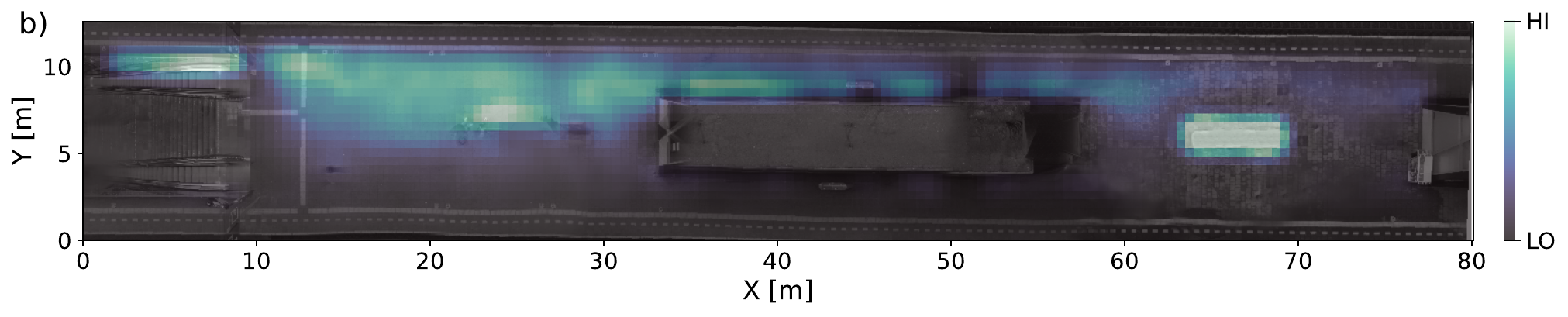}
  \includegraphics[width=\textwidth]{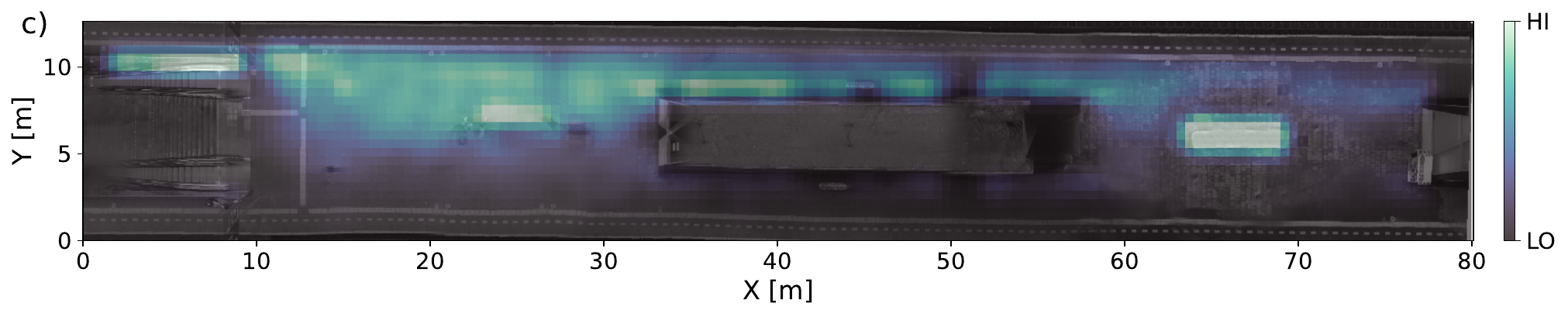}

  \caption{Estimates of the one-particle marginal density $p_1$. (a) Data with $N_{\text{groupmax}}=5$, (b) simulations from DGS process fit on this data, and (c) simulations from DGS process fit on data with macro-pedestrian correction. All are estimated using a Gaussian kernel density estimator with the same bandwidth ($0.5\,$m) for comparability, and are barely distinguishable.}
  \label{fig:heatmaps}
\end{figure}

\section{Results of Semiparametric Statistical Estimation Under Consideration of Social Groups}
\label{sec:results}

In this section, we briefly introduce some metrics of comparison to evaluate how the models perform in capturing the intensity and interaction structure of spatial patterns. We estimate the DGS model's parameters once without group aggregation, i.e. using the dataset with $N_{\text{groupmax}} = 5$, and once on the dataset consisting of macro-pedestrians. We estimate the spatial heterogeneity and the interaction parameters using the same approach as in \cite{Sickert_2026}, restricting ourselves to the same region of the platform described therein. For comparison, we then simulate patterns from each model on the full platform, generating configurations of sizes matching those in the corresponding datasets. The main effect of the macro-pedestrian correction is a larger interaction range $R$, combined with a reduced shape parameter $\alpha$. 

In order to evaluate how well the models capture the intensity of the process, we compare the \emph{one-particle marginal probability densities} of the data and the simulations. For completeness, we define this well-known observable conditionally on $N(\x) = n$ as
\begin{equation}\label{eq:one-particle_marginal}
    p_1^{(n)}(x)
    =
    \int_{\X}\cdots\int_{\X}
    f(x,x_2,\dots,x_n \mid N(\x)=n)\,dx_2\cdots dx_n,
\end{equation}
where $f$ is the pdf defined in \eqref{eq:gibbs_density}. For this work, we consider the one-particle marginal over the aggregated configuration (i.e. the union over all configurations), denoted by $p_1$. 

\begin{figure}[t]
  \centering
  \includegraphics[width=\textwidth]{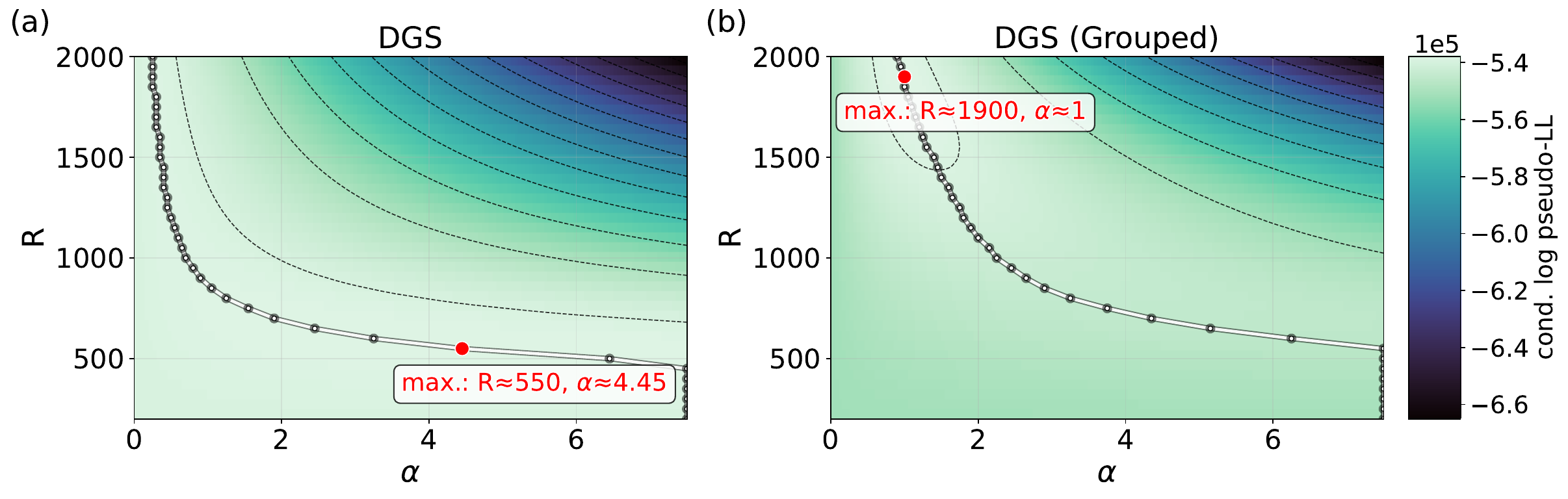}

  \caption{Estimates of the interaction radius $R$ and shape $\alpha$ for DGS model without and with group detection. Considering social groups leads to markedly larger estimated interaction radii, while the shape becomes flatter as $\alpha$ shrinks.}
  \label{fig:parameter_estimates}
\end{figure}

\begin{table}[ht]
  \centering
  \begin{tabular}{lrrr}
    \toprule
     & low occ. $(\leq 30)$ & medium occ. $(31\text{--}60)$ & high occ. $(> 60)$ \\
    \midrule
    original dataset    & 1560 & 669 & 228 \\
    macro-ped. dataset  & 1707   & 600    & 150    \\
    \bottomrule
  \end{tabular}
  \caption{Occupancy split for the original dataset (already restricted to $N_{\text{groupmax}} = 5$) and the macro-pedestrian dataset.}
  \label{tab:occupancy}
\end{table}

To evaluate how effectively each model captures the interaction structure visible in the data, we consider three different distance-based metrics. For this, we separate the data into three groups based on the occupancy level of the platform (a surrogate for crowd density). Occupancy classes are defined heuristically using the number of points after aggregation, i.e. from snapshots containing macro-pedestrians, mirroring the approach in \cite{Sickert_2026}. We consider low occupancy scenarios (with $\leq 30$), medium occupancy (between $31$ and $60$ pedestrians), and high occupancy scenarios ($>60$ pedestrians). The creation of macro-pedestrians skews this distribution in favor of lower-occupancy scenarios, as oftentimes high-occupancy scenarios contain many social groups in the original data, see \tref{tab:occupancy}. The metrics we consider are as follows (see \cite[Section 3]{Sickert_2026} for a more thorough introduction):
\begin{enumerate}[label=(D\arabic*), ref=(D\arabic*)]
    \item \label{item:D1} NN-distance distributions: empirical probability densities of the first and second nearest-neighbor (NN) distances at different occupancy levels should have similar features (peaks and tails should have matching location and behavior).
    \item \label{item:D2} Average ball count $\mathcal{C}(r)$: returns number of points around any given point within a disk of radius $r$, averaged over all points in the snapshot. Formally, for $\{x_1, \dots, x_n\}$ and $r > 0$,
    \[ \mathcal{C}(r) = \frac{1}{n} \sum_{i=1}^n \sum_{k=1}^{n} \1\{ \|x_i - x_k \| \leq r \}, \]
    where $\| \cdot \|$ is Euclidean distance, and $\1$ denotes an indicator function. $\mathcal{C}(r)$ can then be averaged across configurations.
    \item \label{item:D3} Average L-function: this is a modified version of Ripley's K-function~\cite{Ripley_1977}, defined as $L(r) = \sqrt{K(r)/\pi}$, where $K$ is Ripley's K-function, which is a function of the distribution of random configurations $\x \in \X_n^*$, and thus must be estimated. If we assume $\beta$ is homogeneous, we can estimate $K$ by~\cite{Baddeley_2015}
    \begin{equation*}\label{eq:K_hat_hom}
        \hat{K}_{\text{hom}}(r) = \frac{|\X|}{n(n-1)} \sum_{i=1}^n \sum_{\substack{j=1 \\ j \neq i}}^n \1\{\|x_i - x_j \| \leq r \}. 
    \end{equation*}
    This is similar to $\mathcal{C}(r)$, apart from self-counting and normalization. Homogeneity of $\beta$ is a strong assumption, so we may consider an estimator accounting for inhomogeneity instead by inserting an estimate $\hat{\beta}$ of $\beta$ \cite{Baddeley_2015}:
    \begin{equation*}\label{eq:K_hat_inhom}
        \hat{K}_{\text{inhom}}(r) = \frac{1}{|\X|} \sum_{i=1}^n \sum_{\substack{j=1 \\ j \neq i}}^n \frac{1}{\hat{\beta}(x_i) \hat{\beta}(x_j)} \1\{\|x_i - x_j \| \leq r \}.
    \end{equation*}
    If $\beta$ is constant, take $\hat{\beta} = n/|\X|$ to recover $\hat{K}_{\text{hom}}$. An empirical estimate of the L-function (i.e., $\hat{L}_{\text{hom}}$ or $\hat{L}_{\text{inhom}}$) is obtained by plugging in the corresponding K-function estimator. We then take an average value over all configurations.
\end{enumerate}

\begin{figure}[p]
  \centering

  \begin{minipage}[t]{0.32\textwidth}
  \centering
  \includegraphics[width=\textwidth]{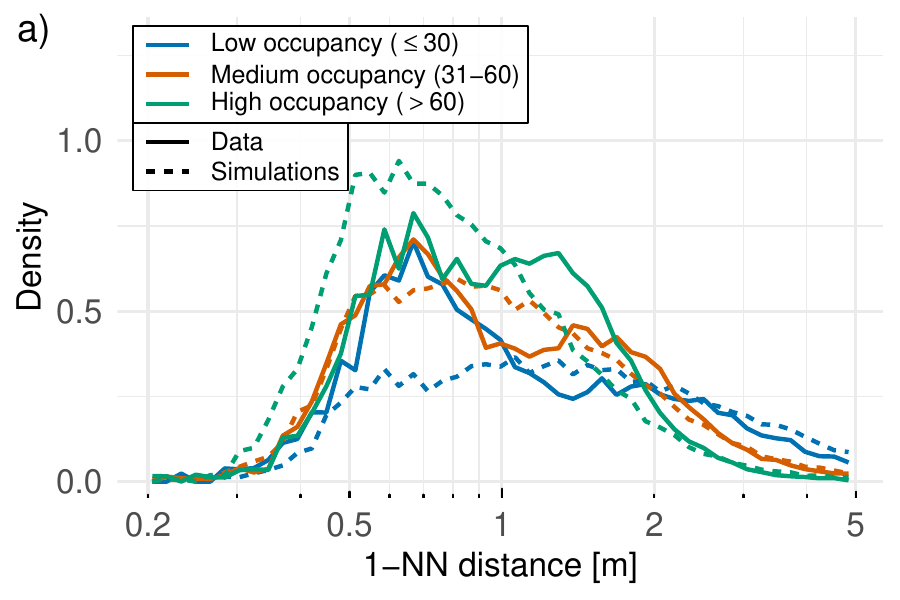}
  \end{minipage}\hfill
  \begin{minipage}[t]{0.32\textwidth}
  \centering
  \includegraphics[width=\textwidth]{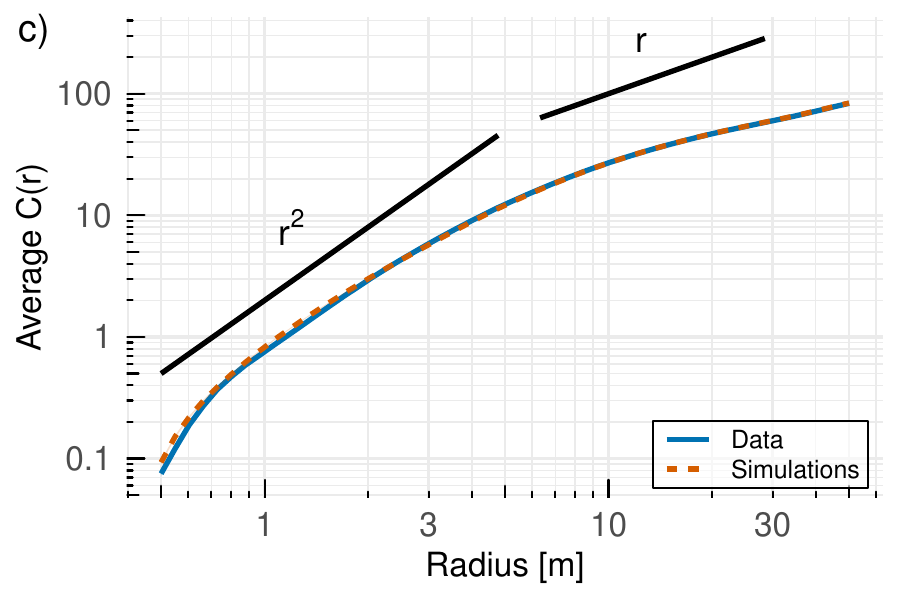}
  \end{minipage}\hfill
  \begin{minipage}[t]{0.32\textwidth}
  \centering
  \includegraphics[width=\textwidth]{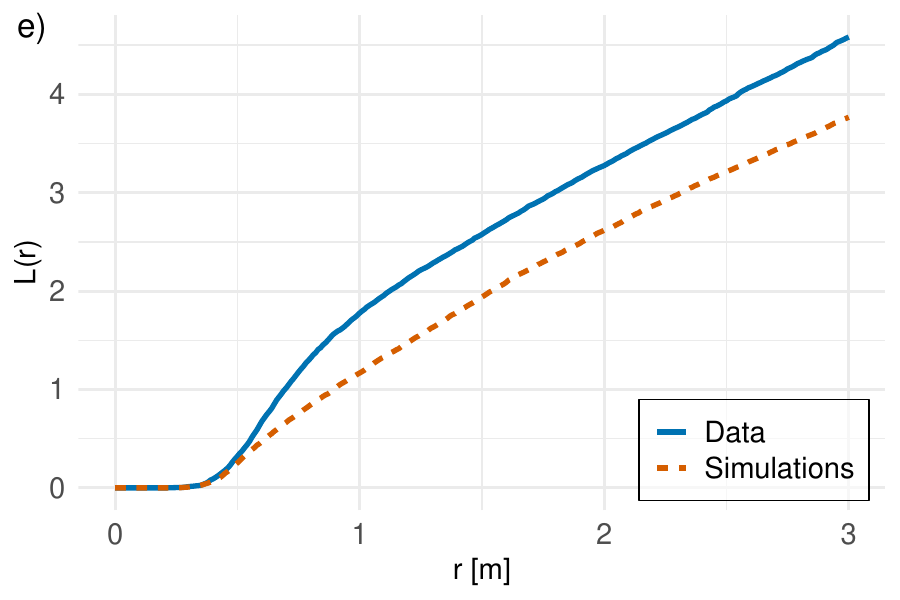}
  \end{minipage}

  \begin{minipage}[t]{0.32\textwidth}
  \centering
  \includegraphics[width=\textwidth]{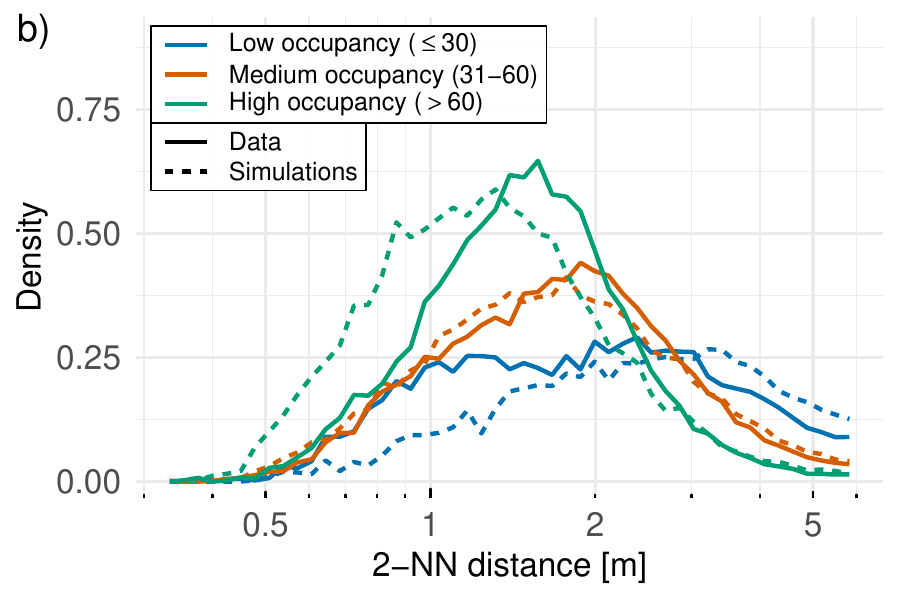}
  \end{minipage}\hfill
  \begin{minipage}[t]{0.32\textwidth}
  \centering
  \includegraphics[width=\textwidth]{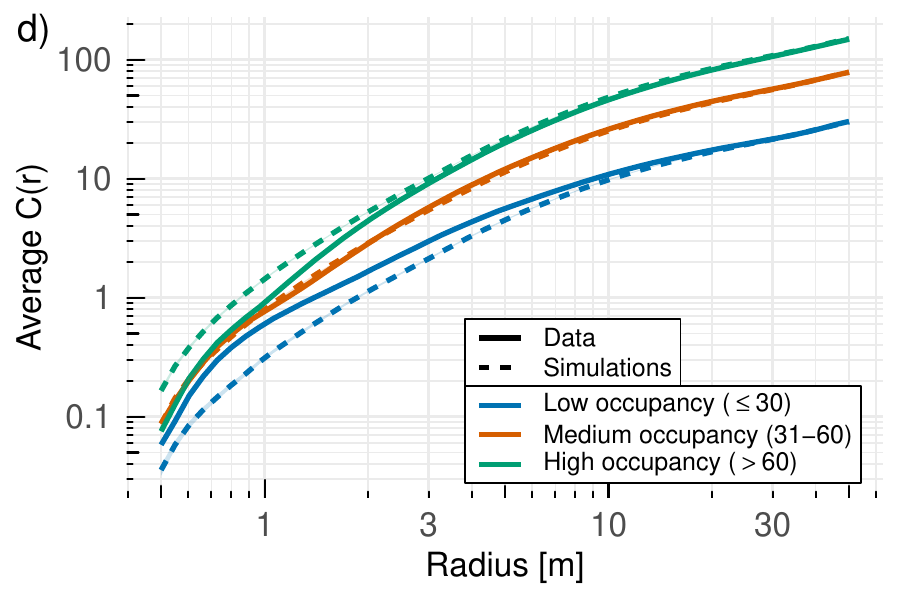}
  \end{minipage}\hfill
  \begin{minipage}[t]{0.32\textwidth}
  \centering
  \includegraphics[width=\textwidth]{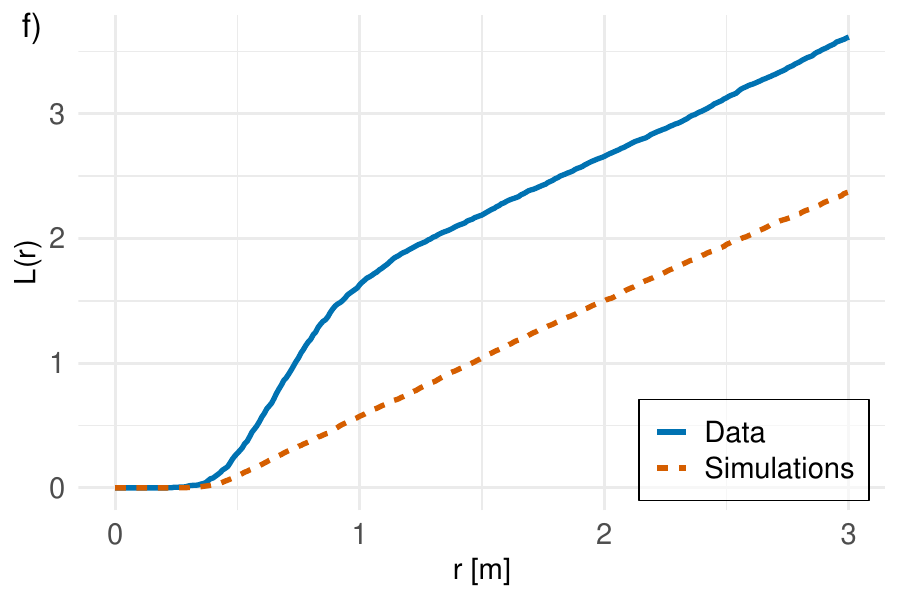}
  \end{minipage}

  \caption{Comparison of distance-based statistics of DGS simulations from model without grouping versus the corresponding data (i.e. with $ N_{\text{groupmax}}=5$). (a) Pdf of NN distances, separated by occupancy. (b) Pdf of 2-NN distances, separated by occupancy. (c) Avg. $\mathcal{C}(r)$ over all snapshots in log-log scale. (d) Avg. $\mathcal{C}(r)$ separated by occupancy. (e) L-functions for simulations vs. data without inhomogeneity correction. (f) L-functions with inhomogeneity correction.}
  \label{fig:nongrouped_dgs}

  \vspace{1em}

  \centering

  \begin{minipage}[t]{0.32\textwidth}
  \centering
  \includegraphics[width=\textwidth]{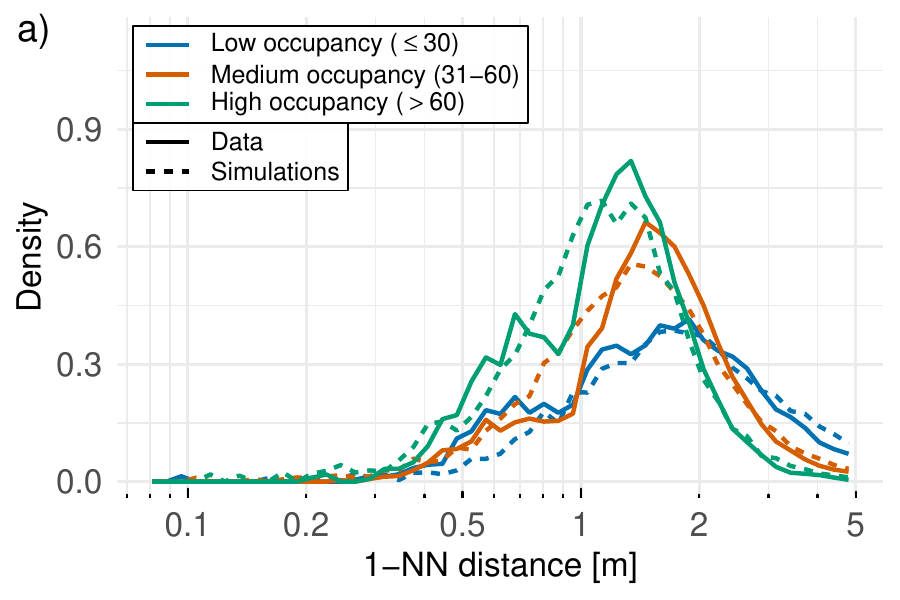}
  \end{minipage}\hfill
  \begin{minipage}[t]{0.32\textwidth}
  \centering
  \includegraphics[width=\textwidth]{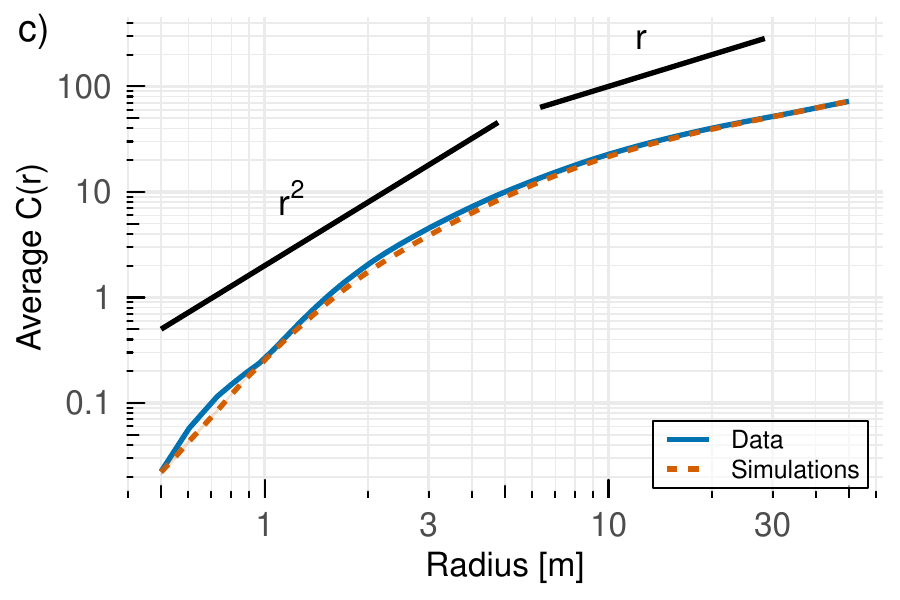}
  \end{minipage}\hfill
  \begin{minipage}[t]{0.32\textwidth}
  \centering
  \includegraphics[width=\textwidth]{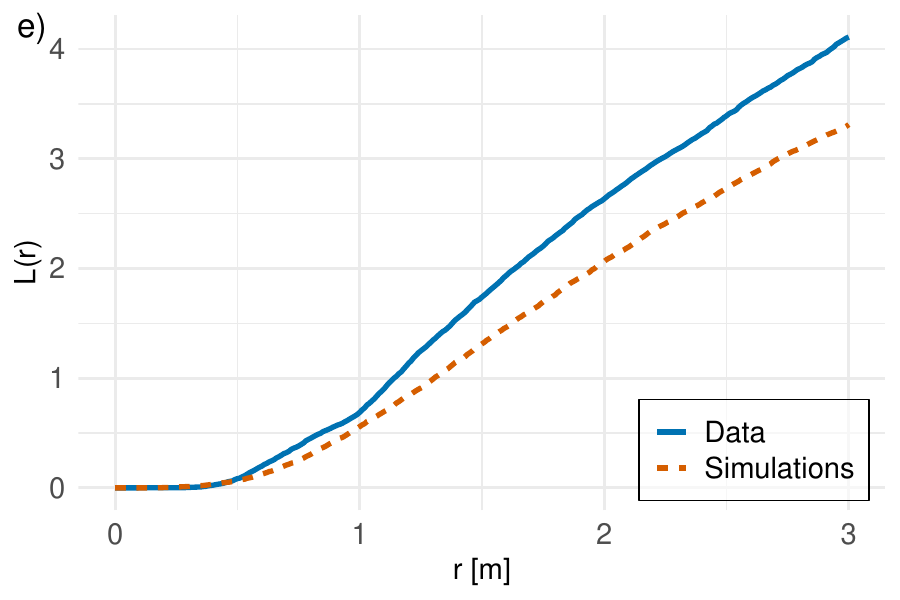}
  \end{minipage}

  \begin{minipage}[t]{0.32\textwidth}
  \centering
  \includegraphics[width=\textwidth]{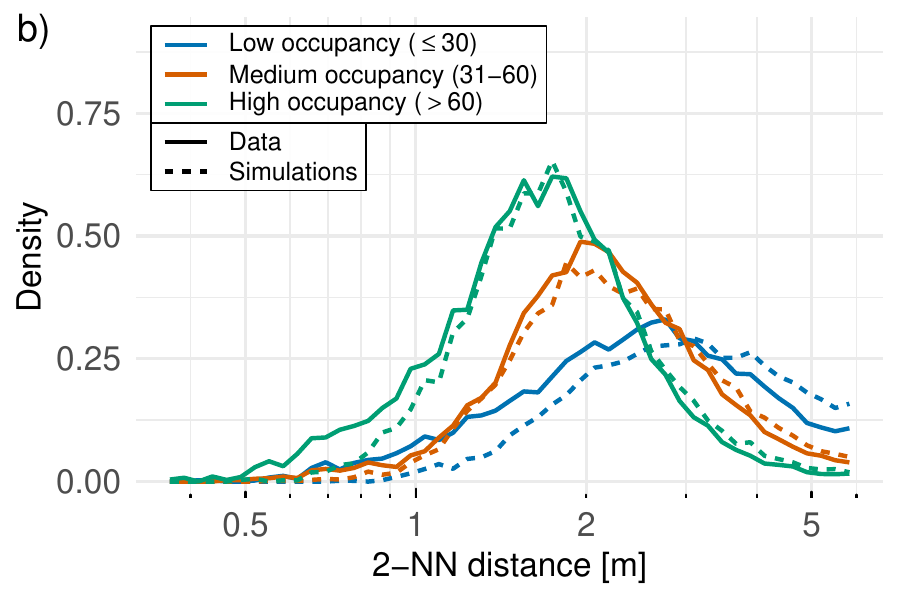}
  \end{minipage}\hfill
  \begin{minipage}[t]{0.32\textwidth}
  \centering
  \includegraphics[width=\textwidth]{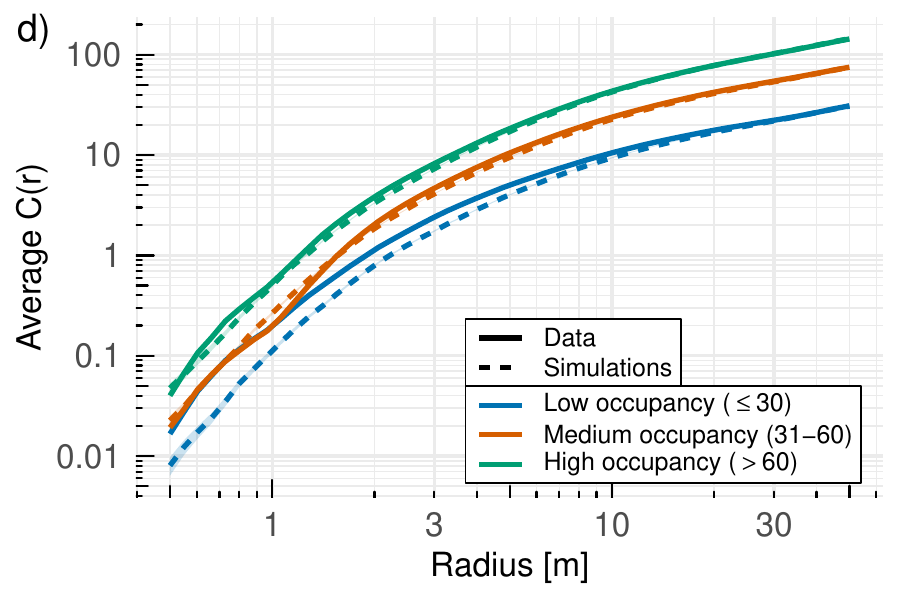}
  \end{minipage}\hfill
  \begin{minipage}[t]{0.32\textwidth}
  \centering
  \includegraphics[width=\textwidth]{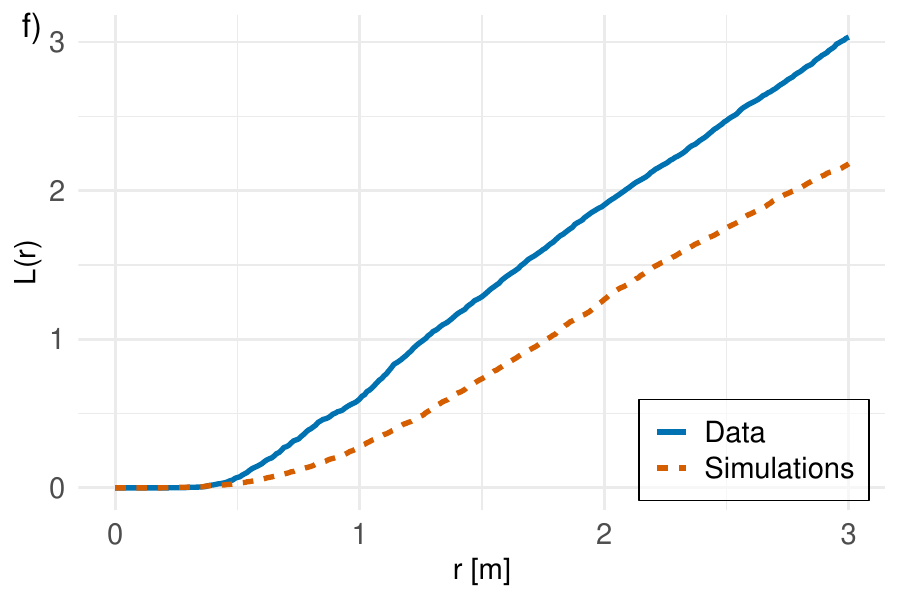}
  \end{minipage}

  \caption{Comparison of distance-based statistics of DGS simulations from the model taking groups of pedestrians into account versus the corresponding data (i.e. snapshots consisting of ``macro-pedestrians''). Comparing with \fref{fig:nongrouped_dgs}, NN distances show a better fit, while performance on other metrics is only slightly improved. (a) Pdf of NN distances, separated by occupancy. (b) Pdf of 2-NN distances, separated by occupancy. (c) Avg. $\mathcal{C}(r)$ over all snapshots in log-log scale. (d) Avg. $\mathcal{C}(r)$ separated by occupancy. (e) L-functions for simulations vs. data (f) L-functions for simulations vs. data with correction for inhomogeneity.}
  \label{fig:grouped_dgs}
\end{figure}

\noindent In terms of the intensity, comparing the one-particle marginal densities $p_1$ seen in \fref{fig:heatmaps} qualitatively shows that there is barely any difference between models. Since the estimate of the intensity function in the Gibbs model is based on $p_1$, high agreement is to be expected across the board. \fref{fig:parameter_estimates} illustrates that the estimates we get for the interaction parameters are visibly different. \fref{fig:interactions} shows the shapes of the resulting interaction functions, compared with parameter estimates obtained without group aggregation. Including social groups increases the radius of interaction $R$, while the shape parameter $\alpha$ shrinks, resulting in a smoother transition between full exclusion and independence. This has varying results on the metrics \ref{item:D1}-\ref{item:D3}, shown in \fref{fig:nongrouped_dgs} for the model without grouping and in \fref{fig:grouped_dgs} for the macro-pedestrian model. For the latter, the NN and 2-NN distance distributions in \fref{fig:grouped_dgs}(a-b) show high agreement at all three occupancy levels, an improvement over the model without grouping, in which a clear peak (likely caused by pairs) is visible at short ranges that is no longer present in the macro-pedestrian dataset (see \fref{fig:nongrouped_dgs}(a-b)). On the other hand, improvements are less clear for the average $\mathcal{C}(r)$ (\fref{fig:grouped_dgs}(c-d)) and the L-functions (\fref{fig:grouped_dgs}(e)). For the average $\mathcal{C}(r)$, values for the simulations are still well below the data at short ranges in low occupancy scenarios, although at medium occupancy agreement is higher than for the model without group aggregation. For the L-function, agreement is higher at short ranges, but the data and simulations still diverge at around $1\,$m. Nonetheless, comparing performance to the model without grouping comes with some caveats. We are no longer comparing two models to the same dataset, as one contains the aggregated macro-pedestrians, while the other does not. Strictly speaking, the evidence only supports the conclusion that the DGS model fits the macro-pedestrian data more closely than it fits the uncorrected data.

Overall, correcting for groups can lead to the model having a better fit than on the uncorrected data. However, the considerable change in the choice of interaction radii indicates that estimates may be biased in the opposite direction from \cite{Sickert_2026} (here, experiments showed that the estimation method tends to underestimate the strength or range of repulsion). The aggregation step removes short within-group distances entirely, replacing multiple pedestrian positions by a single centroid. This changes the geometry of the pattern, as distances from the group to nearby pedestrians are measured from the centroid rather than from the closest group member. Consequently, the new pattern may appear more regular (i.e. repulsive) than the physical pedestrian configuration, which can lead to overestimated repulsion ranges.

\section{Discussion}
\label{sec:discussion}

We examined how social groups affect inference for repulsive spatial point process models of waiting pedestrians and took first steps towards accounting for them. Starting from the model and semiparametric estimation procedure of \cite{Sickert_2026}, we detected groups based on pairwise distances and contact durations, replaced each group by a macro-pedestrian at its centroid, and refitted the modified DGS interaction function to the resulting patterns. This correction yielded a better fit in terms of nearest-neighbor statistics, yet two caveats remain. First, a direct comparison with the group-agnostic model is difficult, as the macro-pedestrian correction changes the underlying data used for estimation and validation. Second, the correction may cause the estimated interaction range to overshoot.

The main result concerns the nearest-neighbor statistics. Simulations under the refitted model reproduce the NN and 2-NN distance distributions of the macro-pedestrian data well at all three occupancy levels, whereas the group-agnostic model shows a pronounced short-range peak, likely caused by pairs. The better fit is accompanied by a considerably larger estimated interaction radius $R$, while the shape parameter $\alpha$ decreases, resulting in a more gradual transition from full exclusion to independence over a longer range. This suggests that proximity within groups accounts for part of the short-range repulsion inferred from the raw data. This improvement must be interpreted with care, since the macro-pedestrian correction changes the data used for estimation and validation. Aggregation reduces the number of points and shifts the occupancy distribution towards lower levels (\tref{tab:occupancy}), and snapshots containing groups of more than five members were excluded altogether, since their centroids can lie far from the group members. Overall, this makes a direct comparison of the goodness of fit difficult.

As mentioned above, we conjecture that positional averaging overcorrects the interaction range $R$. Replacing a group by its centroid removes distance structures within groups entirely, and measures distances to neighboring pedestrians from the center rather than from the nearest group member. The aggregated patterns therefore appear more regular than physical configurations, and the estimation procedure, which tended to underestimate repulsion when groups were ignored \cite{Sickert_2026}, may now be biased in the opposite direction. Positional averaging approach is thus a simple first step to address the influence of groups, but it falls short of a fully adequate treatment.

This opens the door to further, more detailed investigation of modeling social groups in waiting pedestrians. Rather than aggregating the data, social groups could be represented within the model itself, so that all models are fitted to the same individual-level data and can be compared directly. Marked point processes \cite{Baddeley_2015} are promising in this regard, as they allow the interaction to depend on the marks of both points. With marks encoding group membership, pairs within the same group and pairs of strangers could follow separate interaction mechanisms. Alternatively, the macro-pedestrian approach proposed here could be refined by treating macro-pedestrians as disks whose diameters reflect the spatial size of their groups, with interaction depending on the distance between disk boundaries rather than centroids. This could reduce the bias introduced by positional averaging, although representing pedestrians (and groups) as disks remains an approximation which is not necessarily realistic, particularly for large groups.

\begin{contributions}
  The authors confirm contribution to the paper as follows: Conceptualization and Methodology: L. Sickert Karam, R. M. Castro, A. Corbetta; Software, Formal analysis, Investigation, Data curation, Visualization, Writing – original draft and Writing – review \& editing: L. Sickert Karam; Supervision, Writing – review \& editing, Funding acquisition: R. M. Castro, M. Schoukens, A. Corbetta. All authors reviewed the results and approved the final version of the manuscript.
\end{contributions}

\begin{acknowledgements}
  This work was funded by the Eindhoven Artificial Intelligence Systems Institute (EIASI) Exploratory Multidisciplinary Artificial Intelligence Research Program (EMDAIR) project \emph{AICrowd}.
\end{acknowledgements}

\begin{ethics}
  A Large Language Model (Claude Opus 5.5) was used for language editing and debugging code. The authors carefully reviewed all AI-assisted output and take full responsibility for the content of the manuscript.
\end{ethics}

\bibliographystyle{bibstyle}
\bibliography{biblio}

\begin{thebibliography}{10}
\providecommand{\url}[1]{{#1}}
\providecommand{\urlprefix}{URL }
\expandafter\ifx\csname urlstyle\endcsname\relax
  \providecommand{\doi}[1]{DOI~\discretionary{}{}{}#1}\else
  \providecommand{\doi}{DOI~\discretionary{}{}{}\begingroup
  \urlstyle{rm}\Url}\fi

\bibitem{Kupper_2023a}
K{\"u}pper, M., Seyfried, A.: Waiting in crowded places: influence of number of
  pedestrians, waiting time and obstacles.
\newblock Journal of the Royal Society Interface \textbf{20}, 20230193 (2023).
\newblock \doi{10.1098/rsif.2023.0193}

\bibitem{Sickert_2026}
{Sickert Karam}, L., Castro, R.M., Schoukens, M., Corbetta, A.: Modeling
  inhomogeneous spatial point configurations with applications to replicated
  patterns in waiting crowds.
\newblock Pre-print arXiv:2606.14532 [physics.soc-ph] (2026)

\bibitem{Baddeley_2015}
Baddeley, A., Rubak, E., Turner, R.: Spatial Point Patterns: Methodology and
  Applications with R.
\newblock CRC Press (2015)

\bibitem{Helbing_1995}
Helbing, D., Moln{\'a}r, P.: Social force model for pedestrian dynamics.
\newblock Physical Review E \textbf{51}(5), 4282--4286 (1995).
\newblock \doi{10.1103/PhysRevE.51.4282}

\bibitem{vanLieshout_2000}
van Lieshout, M.N.M.: Markov Point Processes and Their Applications.
\newblock Imperial College Press (2000).
\newblock \doi{10.1142/p060}

\bibitem{Diggle_1987}
Diggle, P.J., Gates, D.J., Stibbard, A.: A nonparametric estimator for
  pairwise-interaction point processes.
\newblock Biometrika \textbf{74}(4), 763--770 (1987).
\newblock \doi{10.2307/2336470}

\bibitem{Kupper_2023b}
K{\"u}pper, M., Seyfried, A.: Identification of social groups and waiting
  pedestrians at railway platforms using trajectory data.
\newblock PLOS ONE \textbf{18}, 1--17 (2023).
\newblock \doi{10.1371/journal.pone.0282526}

\bibitem{Sickert_2026b}
{Sickert Karam}, L., Castro, R.M., Schoukens, M., Corbetta, A.: Modeling
  inhomogeneous spatial point configurations with applications to replicated
  patterns in waiting crowds - dataset: snapshots of pedestrians at eindhoven
  train station (2026).
\newblock \doi{10.5281/zenodo.23060582}.
\newblock Dataset

\bibitem{Pouw_2020}
Pouw, C.A.S., Toschi, F., van Schadewijk, F., Corbetta, A.: Monitoring physical
  distancing for crowd management: Real-time trajectory and group analysis.
\newblock PLOS ONE \textbf{15}(10), 1--18 (2020).
\newblock \doi{10.1371/journal.pone.0240963}

\bibitem{vanderLaan_2026}
van~der Laan, C., Corbetta, A.: Probabilistic dynamics of small groups in crowd
  flows.
\newblock PLOS ONE \textbf{21}, 1--26 (2026).
\newblock \doi{10.1371/journal.pone.0356173}

\bibitem{Geyer_1994}
Geyer, C.J., M{\o}ller, J.: Simulation procedures and likelihood inference for
  spatial point processes.
\newblock Scandinavian Journal of Statistics \textbf{21}(4), 359--373 (1994)

\bibitem{Besag_1975}
Besag, J.: Statistical analysis of non-lattice data.
\newblock Journal of the Royal Statistical Society: Series D (The Statistician)
  \textbf{24}(3), 179--195 (1975).
\newblock \doi{10.2307/2987782}

\bibitem{Moller_2022}
M{\o}ller, J., Vihrs, N.: Should we condition on the number of points when
  modelling spatial point patterns?
\newblock International Statistical Review \textbf{90}, 551--562 (2022).
\newblock \doi{10.1111/insr.12501}

\bibitem{Ripley_1977}
Ripley, B.D.: Modelling spatial patterns.
\newblock Journal of the Royal Statistical Society. Series B (Methodological)
  \textbf{39}(2), 172--212 (1977).
\newblock \doi{10.1111/j.2517-6161.1977.tb01615.x}

\end{thebibliography}

\end{document}